# Perfect and Stable Hybrid Glasses from Strong and Fragile Metal-Organic Framework Liquids

**Authors:** T. D. Bennett[1]†, J. C. Tan[2]†, Y. Z. Yue†[3,4], C. Ducati[1], N. Terril[5], H. H. M. Yeung[6], Z. Zhou[7], W. Chen[7], S. Henke[1], A. K. Cheetham[1] and G. N. Greaves[1, 4, 7]†*

**Affiliations:**

[1] Department of Materials Science and Metallurgy, University of Cambridge, Charles Babbage Road, Cambridge, CB2 3QZ, United Kingdom.

[2] Department of Engineering Science, University of Oxford, Parks Road, OX1 3PJ, Oxford, United Kingdom.

[3] Section of Chemistry, Aalborg University, DK-9220 Aalborg, Denmark

[4] State Key Laboratory of Silicate Materials for Architectures, Wuhan University of Technology, Wuhan 430070, China

[5] Diamond Light Source Ltd, Diamond House, Harwell Science and Innovation Campus, Didcot, OX11 0DE, United Kingdom.

[6] International Center of Materials Nanoarchitectonics (MANA), National Institute for Materials Science (NIMS), Namiki 1-1, Tsukuba, Ibaraki 305-0044, Japan.

[7] Institute of Mathematics and Physics, Aberystwyth University, Aberystwyth, SY23 3BZ, United Kingdom.

*To whom correspondence should be addressed. Email gng25@cam.ac.uk.

†These authors contributed equally to this work.

Hybrid glasses connect emerging fields of metal-organic frameworks (MOFs) with the glass-formation, amorphization, and melting processes of these structurally diverse and chemically versatile systems[1-3]. Most zeolites, including MOFs, amorphize around the glass transition, devitrifying and then melting at much higher temperatures[4-15]. The relationship between the two processes has so far not been investigated. Herein we show how heating first results in a low density 'perfect' glass, following an order-order transition, leading to a super-strong liquid of low fragility that dynamically controls MOF collapse. A subsequent order-disorder transition creates a high density liquid of greater fragility. After crystallization and melting, subsequent cooling results in a stable glass virtually identical to the high density phase. Furthermore, the wide-ranging melting temperatures of different MOFs suggest these can be differentiated by topology. Our research provides new insight into the stability and functionality of these novel ductile crystalline materials, including the possibility of 'melt-casting' MOFs.

Microporous hybrid materials, known as metal-organic frameworks (MOFs), consist of inorganic clusters or ions bridged by organic ligands in open three-dimensional arrays, and are of great interest due to their potential use in gas separation and storage, drug delivery, catalysis and sensing applications[1-3]. An important subset of MOFs, the zeolitic imidazolate frameworks (ZIFs) display some striking similarities to zeolites (purely inorganic porous frameworks of corner sharing $SiO_4$ and $AlO_4$ tetrahedra), adopting similar network structures[4] and, in particular, undergoing time-dependent thermal and pressure-induced amorphization (loss of periodicity)[5-7]. High-density amorphous (HDA) inorganic glasses, along with low density amorphous (LDA) states of identical topologies and vibrational entropies to their parent crystalline phases, have previously been produced via zeolite amorphization [5,7]. The latter, which we term 'perfect' glasses, are of particular scientific interest due to their location deep in the potential energy landscape (PEL) [8], and their unique mechanical properties [9,10]. Ultrastable nanostructured polymer glasses have also recently been produced by molecule-by-molecule deposition [11], offering a facile route to low density thermally stable films, though are reliant on a preponderance of Voronoi polyhedra [12] and possess limited chemical functionality. The present work, however, explores the amorphization of the more versatile MOFs, revealing a complex sequence of events that is unprecedented in other glass-forming systems. We concentrate on ZIF-4 $[Zn(C_3H_3N_2)_2]$, contrasting this with ZIF-8 $[Zn(C_4H_5N_2)_2]$ [13]. Both crystalline structures comprise even-membered rings which are retained on amorphization (Fig. 1). We have discovered that heating ZIF-4, to ~600 K yields an HDA glass of identical composition via a LDA 'perfect' glass phase. [14] On further heating to ~700 K, the high density HDL super-cooled liquid recrystallizes to a dense crystalline framework, ZIF-zni with a melting point $T_m$ ~850 K, before decomposing at ~900 K (Fig. 1).

Herein we study the mechanism of amorphization in much greater detail by thermogravimetric analysis and heat capacity measurements (Fig. 1), as well as by *in situ* SAXS/WAXS experiments (Fig. 2). The loss of solvent at ~550 K is followed by collapse of the crystalline framework to LDA, with a glass transition temperature $T_g$ at 589 K (Fig. 3), characterized by the formation of an LDL super-cooled liquid. The latter phase has very low fragility, $m$=14 that is typical of a super-strong glass-former[8,14]. The LDL super-cooled liquid then coverts at 638 K through a order-disorder transition to an HDL phase with much high fragility ($m$=41) (Fig. 4). The two "polyamorphic" phases (LDL and HDL) coexist in the temperature range from around 635 K to 655 K and the mixture proves to be mechanically auxetic (negative Poisson's ratio[15]) (Fig. 4). Quenching the HDL liquid produces an HDA amorphous phase whose $T_g$ is 565 K (Fig. 3). Importantly, the different $T_g$'s of LDA and HDA reflect their differences in fragility and their differing depths in the PEL (Fig. 4). In addition, quenching the ZIF-zni melt from below the dissociation temperature leads to the formation of a bulk glass with virtually the same composition and $T_g$ as the HDA phase. Moreover $T_m$ for ZIF-4 lies close to that of an inorganic phosphate with a related zeolitic topology[16]. ZIF-8, on the other hand, adopts the sodalite structure, and while it amorphizes under pressure[17], it decomposes before it melts, but a "virtual" $T_m$ can be calculated, lying close to the "real" $T_m$ of its inorganic counterpart (Fig. 4). This suggests the dominance of network architecture in melting, characterized by collective terahertz (THz) vibrations[7], in contrast to the simplistic interatomic variance condition for melting enshrined in Lindemann's Law [18].

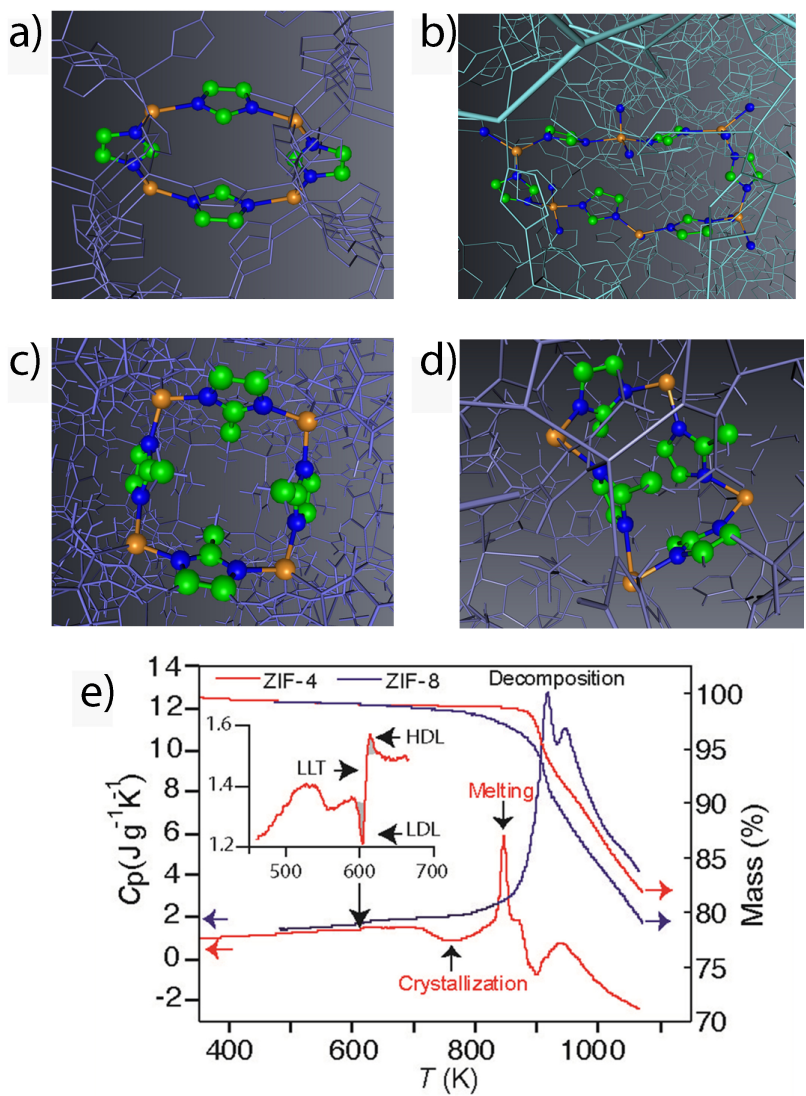

**Fig. 1.** Visualization of ZIF topology, Thermogravimetric analysis (TGA) and heat capacity ($C_p$). **(a)** Highlighting the rings common in zeolitic topologies, in the ordered structure of crystalline ZIF-4. Zn – orange, N – blue, C – green, H atoms omitted for clarity. **(b)** The disordered HDA phase gained via reverse Monte-Carlo modeling [19]. **(c)** Crystalline ZIF-8, which does not amorphize upon heating. **(d)** Pressure-induced contorted HDA topology of ZIF-8. **(e)** TGA and associated $C_p$ plots for both ZIF-4 and ZIF-8, showing, for the former (inset), the collapse to LDL phase (0.75 J g$^{-1}$, exothermic) which is closely followed by formation of the HDL phase (0.44 J g$^{-1}$, endothermic), and recrystallization into the dense ZIF-zni (exothermic). Heating rates were 20 K min$^{-1}$ for the main traces and 10 K min$^{-1}$ for the inset. Melting (endothermic) then follows before thermal degradation.

Variable temperature small- and wide-angle X-ray scattering (SAXS/WAXS) measurements were performed to probe the mechanism of amorphization. The SAXS invariant [8] $Q$ (which measures differences in local density) crucially increases after the disappearance of Bragg diffraction from the sample at 634 K, and remains at a maximum to 653 K (Fig. 2 a, b and Fig. S1), supporting the coexistence of two polyamorphic liquid phases (LDL and HDL) for this hybrid system, identical in composition but different in density. In contrast, the structural integrity of ZIF-8 was maintained throughout the heating and cooling cycle (Fig. 2 c, d).

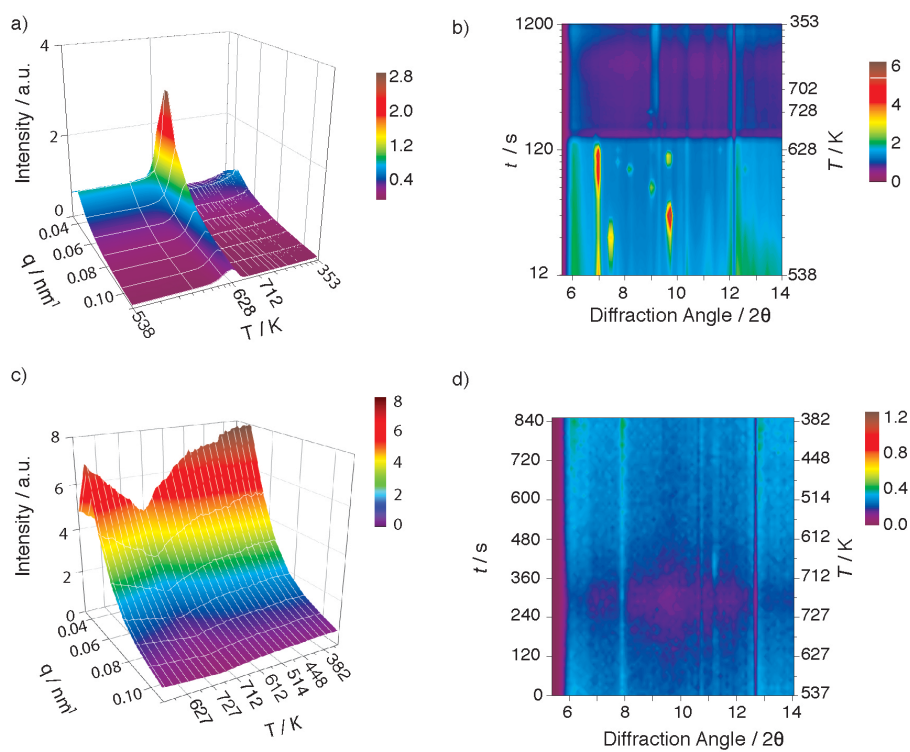

**Fig. 2.** *In-situ* SAXS/WAXS data collected on ZIF-4 (top) and ZIF-8 (bottom) **(a)** 3D plot of $Q_{SAXS}$ of ZIF-4, clearly indicates the emergence of a peak between 634 and 653 K. **(b)** WAXS data shows the loss of Bragg diffraction on collapse at *ca.* 630 K. Subsequent diffraction lines due to the ensuing crystallization to ZIF-zni. **(c)** 3D plot of the SAXS results for ZIF-8, the dip in intensity relates to the reverse in heating direction. **(d)** WAXS data shows the retention of crystallinity across the temperature range studied. Heating rate 50 K min$^{-1}$.

Differential scanning calorimetry (DSC) measurements on ZIF-4 contain endothermic features attributed to the release of pore-occupying molecules (the framework template N, N – dimethylformamide) at ~550 K, before an exothermic feature starting at 589 K, indicating the LDA $T_g$ (confirmed by SAXS experiments, Fig. 3a) anticipating the collapse of the crystalline framework at ~630 K (Fig. 3a). The order-order ZIF-4→LDA transition is followed by an endothermic liquid-liquid LDL-HDL order-disorder transition, complete by 638 K, similar to polyamorphic transitions in zeolites [5,14] and glass-forming liquids[8,15]. The transition is obscured at higher heating rates (Fig. S2). After cooling back to room temperature, the absence of exothermic features on rescanning confirms retention of solvent-free HDA upon reheating (Fig. 3b). $T_g$ for HDA (565 K) is significantly lower than LDA (589 K), reflecting the striking differences in fragility ($m$ = 41, 14 respectively) (Figs. 3a and 4a).

The HDL phase recrystallizes into a dense hybrid framework (ZIF-zni) with an exotherm at ~700 K [19]. On further heating, combined DSC-TGA experiments reveal a sharp melting endotherm at ~ 850 K, neither accompanied by a decrease in framework mass (Fig. 1); both enthalpy changes are identical (Fig. S2b). Investigating the recovered MOF-glass at room temperature using diffraction, spectroscopy and elemental analysis (SI) confirmed the composition being very close to $Zn(C_3H_3N_2)_2$, with almost identical local vibrational modes as crystalline ZIF-4 (Figs. S3, S4). Intriguingly, macroscopic flow of the melt into a glass can be seen in microscopy images (Figs. 3d and S5). From the separation of the striations and thickness of the glass we estimate its yield strength, $\sigma_Y$, to be ~250 MPa (SI), which is typical of hard brittle materials.

ZIF-4 also collapses with pressure at room temperature (RT) between 0.35 and 0.98 GPa [6], equivalent to thermal amorphization between 603 K and 638 K. Following earlier work on zeolite instability [5,14], thermobaric parameters for ZIF-4 collapse result in the T-P phase diagram shown in Fig. 4b. Together with zeolites, ZIF-4 exhibits a critical point C at negative pressure, leading to a sharp increase in density fluctuations around $T_c$ (659 K) and $P_c$ (–0.063 GPa), which, with extension to ambient pressure, explains the sharp $Q_{SAXS}$ peak (Fig. 2).

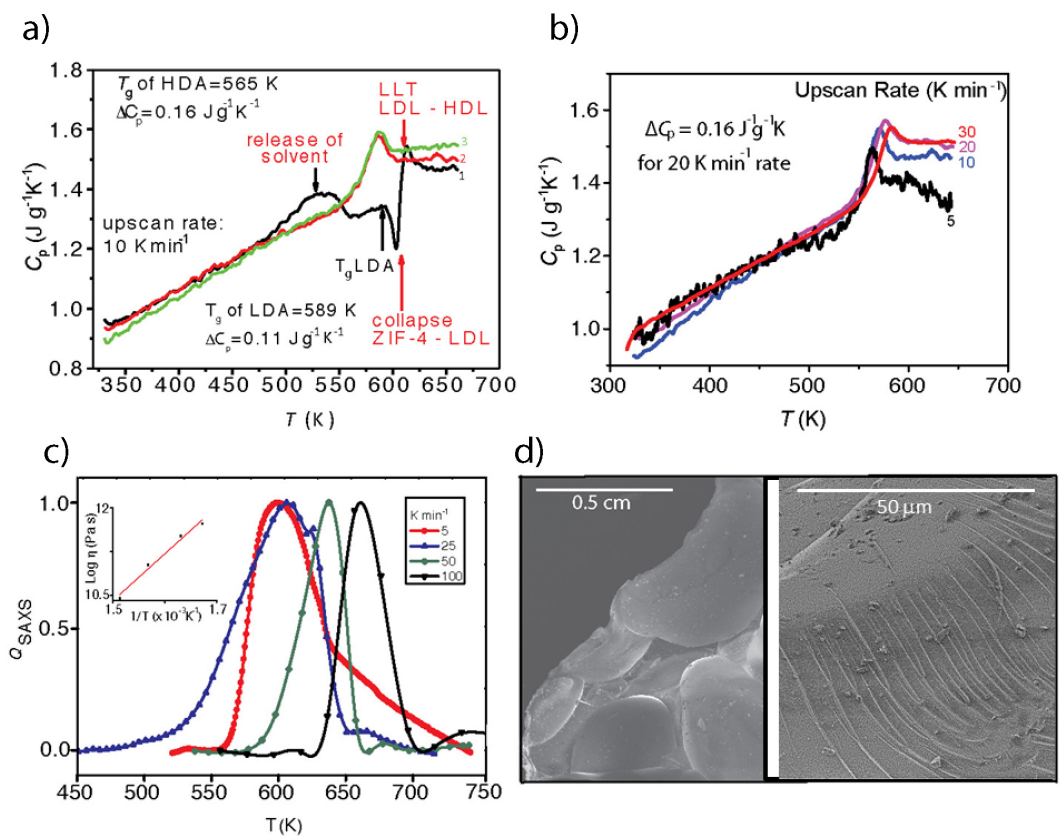

**Fig. 3**. Dynamics of ZIF-4 amorphization, melting and quenching. (**a**) Sequence of DSC upscans on ZIF-4 at 10 K min$^{-1}$ starting with ZIF-4 (1,black), showing: solvent release, collapse to LDA, followed by LDL-HDL transition. $\Delta C_P$ values measured for LDA and HDA phases at $T_g$ are shown. Endotherms in successive scans (2 ,red, 3, green) relate to HDA phase. (**b**) DSC second upscans on the same samples at different rates right after cooling, yielding $T_g$ and $m$ for HDA. (**c**) The change in $Q_{SAXS}$ for different

heating rates, showing the increase of the peak temperature ($T_{peak}$) with increasing heating rates, giving $T_g$ and $m$ for the LDA phase. **Inset:** Dependences of the reciprocal heating rate ($1/q_h$) on the $T_g$-scaled peak temperature ($T_{peak}$). The Maxwell viscosity [8] $\eta = G_\infty \tau$, where $G_\infty$ and $\tau \sim 1/q_h$ are adiabatic shear modulus and structural relaxation time, respectively. **(d)** Scanning electron micrograph images of the recovered MOF glass showing thermal stress-induced striations.

The large differences in viscosity of LDL and HDL phases can be quantified via Angell plots (log $\eta$ versus $T_g/T$) of these two glass-forming liquids (Fig. 4a), with respective fragilities of $m = 14$ and $41$, for the LDL and HDL phases resulting from use of structural relaxation times in SAXS and DSC experiments (SI). Arrhenius ZIF-4→LDL collapse is hence what is expected for very strong liquids ($m = 14$), while the latter ($m = 41$) for HDL has intermediate fragility (Fig. 4a), in comparison to silica which is strong, the relatively fragile anorthite and the very fragile triphenylethene. Comparison of the melt fragility of the LDL with silica ($m = 20$), reveals the initial ZIF-4 liquid to be amongst the strongest yet discovered, and hence is referred to as a super-strong liquid. LDL→HDL (Fig. 1e) is therefore a clear example of a strong-to-fragile transition upon heating [20].

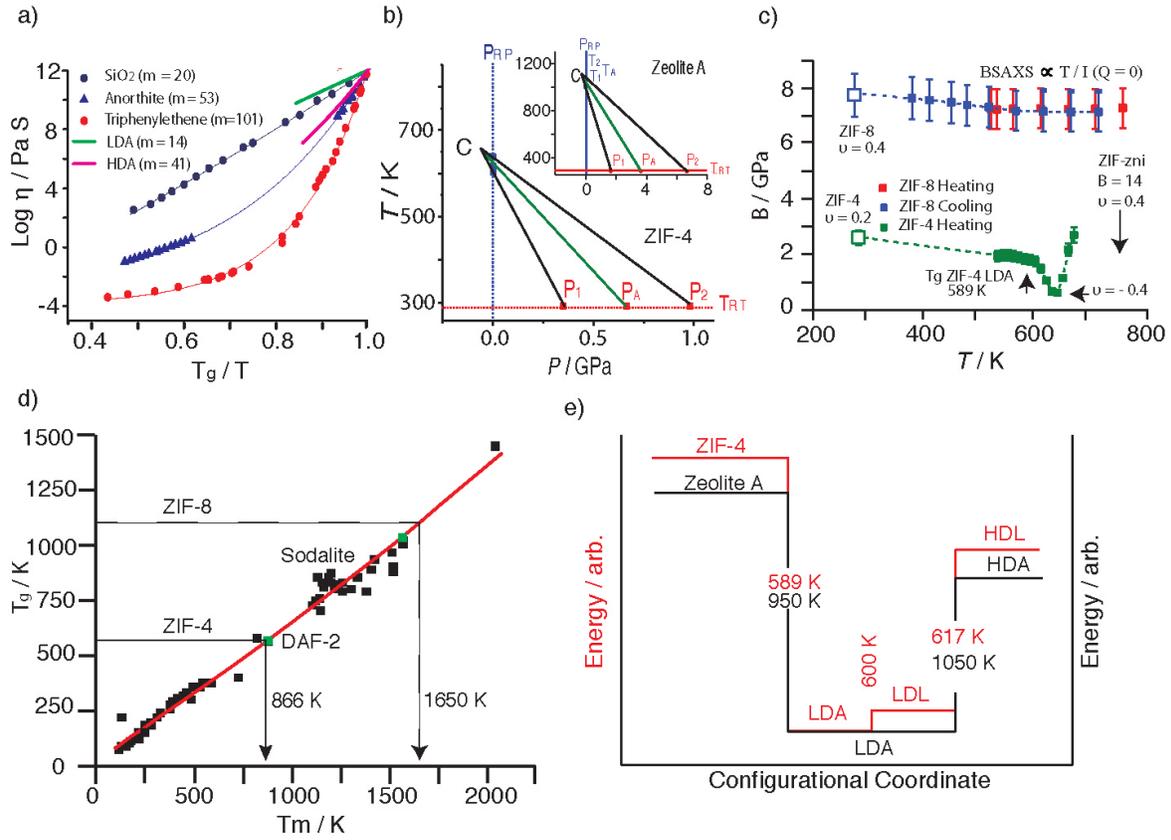

**Fig. 4.** LDL→HDL viscosities, fragilities, critical point, and auxetic behavior; compatibility of amorphization and melting with 2/3's Law and PEL representation. **(a)** Angell plot showing the fragility of LDL and HDL ZIF-4, alongside other glass-forming liquids [21]. Fragility, *m*, is measured from $m = \left(\frac{d(\log\eta)}{d(T_g/T)}\right)_{T=T_g}$. Solid lines are fits to the measured viscosity-temperature relation of the model derived in Ref. 21. **(b)** T-P phase diagrams obtained from thermobaric amorphization parameters for ZIF-4, compared to Zeolite-A (inset [8]). **(c)** Evolution of bulk modulus for ZIF-4 and ZIF-8 as a function of temperature, using a heating rate of 50 K min$^{-1}$. Bulk moduli at RT taken from existing literature [6,22], using constant shear moduli (SI). **(d)** 2/3's Law ($T_g$ v $T_m$) for different glass-forming systems [23-26], including ZIF-4 and ZIF-8 compared to DAF-2 and sodalite respectively (*16,29*). **(E)** PEL for ZIF-4 compared to zeolite-A (*14*) obtained from DSC experiments. The steps are proportional to the exothermic (ZIF-4-LDA) and endothermic (LDL-HDL) transitions from Fig. 1.

Previous theoretical determination of the adiabatic shear modulus ($G_\infty$), along with the relationship $\kappa \propto I(Q=0)/T$, facilitates *in situ* monitoring of the isothermal compressibility $\kappa$ and Poisson's ratio $\nu$ [15] – here for ZIF-4 amorphization (Fig. 4c, SI). The bulk modulus $B=1/\kappa$ decreases rapidly through the collapse of ZIF-4 to the LDA phase, reaching a minimum (where $\nu_{min}$ = -0.4), before recovering through the LDL–HDL transition. Characterization of the coexistence of the LDL-HDL phases as a dilational auxetic material ($\nu<0$), is consistent with modulation of density fluctuations in critical fluids during phase transitions[15], and the associated peak in compressibility $\kappa$ in the vicinity of critical points C[27].

The LDA glass transition temperature $T_g$ (589 K) is extremely close to 2/3 $T_m$ for ZIF=4 (866 K) and therefore complies with the empirical 2/3's law[23] (Fig. 4d). Interestingly, another MOF of Zn(Im)$_2$ composition (ZIF-3), possesses the 'dft' zeolitic topology and undergoes amorphization and recrystallization at identical temperatures to that of ZIF-4 (which adopts the variscite mineral topology, cag) [4]. The inorganic cobalt phosphate framework DAF-2 (also adopting the DFT topology) is observed to melt at 873 K [16], indicating that frameworks with similar network topologies may exhibit similar melting behavior, driven by collective THz modes [7,15]. In contrast, for Debye solids like dense minerals and metals, melting is activated by nearest neighbor $r_{NN}$ vibrations when $\sqrt{(\Delta r^2_{NN})}/r_{NN} \geq 0.1$ - Lindemann's Law [18].

The 2/3's Law, already demonstrated by molecular[24,25] and network glass formers [26], enables $T_m$ to be projected from the LDA $T_g$, where zeolites and MOFs collapse. In particular, despite ZIF-8 undergoing thermal decomposition before melting (Fig. 1e), a 'hypothetical' $T_g$ of 1102 K, can be calculated from the relationship $P_a \Delta V \approx 3RT_g$, where $P_a$ is the amorphization pressure [5,6,17,28].

Using Fig. 4d this projects a 'virtual' $T_m$ for ZIF-8 at 1650 K. The temperature, which in practice is not achieved before dissociation of the hybrid framework, lies close to $T_m$ of the inorganic analogue (sodalite), melting at 1557 K [29,30]. We postulate that, by this methodology, comparison of MOFs with their inorganic analogues should reveal candidates with achievable melting points.

Fundamental to current understanding of the 2/3's Law [24,25] is melt fragility $m$ and its association with the step in specific heat at $T_g$, $\Delta C_P(T_g)$, and the heat of fusion, $H_m$. Recognizing the relationship between melt fragility and Poisson's ratio for the glass [30], we have adapted this empirical relationship: viz, $\Delta C_P(T_g) = \frac{m.H_m}{56 T_g.v}$, and recovered $\Delta C_P$ values measured for LDL and HDL phases (Figs. 3a, SI). Finally, taking the enthalpy changes for ZIF-4 → LDL and LDL→ HDL (Fig. 1), the PEL for ZIF-4, LDL and HDL is plotted schematically, showing strong similarity with the corresponding region for zeolite-A and its polyamorphs (Fig. 4e) [14].

Comparisons between amorphization conditions of MOFs and inorganics may provide further routes to more functional 'perfect' glasses and stable HDA phases. Furthermore, the *in-situ* hybrid liquid formation discovered here opens up possibilities for liquid casting and shaping MOFs into a variety of different solid forms, promising to be an extremely exciting step forward in producing chemically functionalizable hybrid glass materials. The work in its entirety represents a first foray into the new field of MOF glasses.

# Methods

**Synthesis.** The synthesis of ZIF-4 was performed by a modified procedure based on synthesismethods reported in the literature [6,13]. 1.2 g of $Zn(NO_3)_2 \cdot 6H_2O$ and 0.9 g of imidazolewere dissolved in 90 mL of N,N-dimethylformamide (DMF) and transferred into a100 mL screw jar. The jar was tightly sealed and heated to 100 °C for 72 h in an oven.After cooling to room temperature colourless block-shaped crystals were filtered off and first washed three times with ~30 mL pure DMF and then three times with ~30 mL $CH_2Cl_2$. The crystals were gently stirred in 100 mL fresh $CH_2Cl_2$ overnight. Afterwards the solid material was filtered off, washed again three times with ~30 mL fresh $CH_2Cl_2$ and dried in vacuo at 130 °C, using a vacuum oven to yield activated guest-free ZIF-4.

**Measurements.** Temperature dependent *in situ* small angle X-ray scattering (SAXS) and wide angle X-ray scattering (WAXS) measurements were performed on Beamline I22 at the Diamond Light Source synchrotron in the Rutherford Appleton Laboratory (Didcot, Oxfordshire, UK). Detector calibrations were carried out using silver behenate and NBS silicon standards on the RAPID 2D SAXS detector and the HOTWAXS 1D WAXS detector, respectively[31,32]. Scattering data were recorded at 1 Å for angular range of up to 1° for SAXS and over a $2\theta$ range of 5° to 40° for WAXS. Powdered samples of ZIF-4 and -8 were loaded in glass capillaries and inserted horizontally through the Linkam furnace, which was positioned across the synchrotron radiation source.

The apparent isobaric heat capacity ($C_p$) of each sample was measured using a Netzsch STA 449C differential scanning calorimeter (DSC). The samples were placed in a platinum crucible situated on a sample holder of the DSC at room temperature and subjected to varying numbers of up- and –down scans, depending on the purpose of the measurements. After natural cooling to room temperature, the subsequent up-scans were performed using the same procedure as for the first.

FT-IR experiments were performed using a Bruker Tensor 27 Infrared spectrometer, and show retention of interatomic vibrational modes of ZIF-4 in the melt-quenched MOF glass (Fig. S3).

Powder X-ray diffraction measurements on evacuated, guest free ZIF-4 were recorded on a well ground sample with a Bruker D8 Advance powder diffractometer using Cu$K\alpha$ radiation ($\lambda$ = 1.5418 Å) and a LynxEye position sensitive detector in Bragg-Brentano ($\theta-\theta$) geometry at room temperature. Pawley fit shown in Fig. S4a.

Microanalysis was performed at the Department of Chemistry, University of Cambridge as a technical service.
*ZIF-4 Evacuated:* Calculated (based on $Zn(C_3H_3N_2)_2$ composition): C 36.18 %, H 3.02 %, N 28.14 %. Found: C 36.22 %, H 2.98 %, N 28.09 %
*MOF Glass:* Calculated (based on $Zn(C_3H_3N_2)_2$ composition): C 36.18 %, H 3.02 %, N 28.14 %. Found: C 35.64 %, H 2.90 %, N 26.46 %

Optical images of ZIF-4, ZIF-zni and recovered melt-quenched glass are shown in Fig. S5. Scanning Electron Microscope images (Fig. 3D) were taken with an FEI Nova NanSEM (field emission gun). Specimens for SEM analysis were prepared by dispersing fragments of the ZIF-4 melt-quenched glass on conductive carbon tabs for topographic contrast imaging.

**Acknowledgments:**

Data available in the supplementary materials section. The authors would like to thank Trinity Hall (TDB); HRH Sheikh Saud Bin Saqr Al Qasimi (TDB and AKC); Wuhan University of Science and Technology (YY and GNG), ERC grant number 259619 PHOTO EM (CD; Alexander von Humbolt Foundation (SH); Research Center Initiative on Materials Nanoarchitectronics (WAS-MANA) from MEXT, Japan (HHMY).We acknowledge the provision of synchrotron access to Beamline I22 (exp. SM5692) at the Diamond Light Source, Rutherford Appleton Laboratory U.K.

**Author Contributions:**

G.N.G. with A.K.C. facilitating MOF materials and characterization. T.D.B., Y.Y. and G.N.G. wrote the manuscript, J.C.T., with A.K.C. involved in refinements. J.C.T., H.H.M.Y., N.T and G.N.G. carried out the *in sit*u SAXS and WAXS experiments at the Diamond Light Source; JCT analyzed the synchrotron data and established the melt fragilities guided by G.N.G. Y.Y. performed the DSC measurements and data analysis. S.H. synthesized and characterized ZIF-4 crystals for DSC and melting experiments; T.D.B. carried out PXRD, SDT and FTIR measurements on ZIF-4, elemental analysis and optical microscopy on recovered MOF glass. C.D. performed SEM imaging of MOF glass. W.C and Z.Z contributed to- visualization and graphics, including modeling amorphized ZIF-8. J.C.T conducted elasticity and stress analyses. T.D.B. and S.H. involved in the preparation of powder samples used in this study.

**Competing Financial Interests:** The authors have no competing financial interests.